\documentclass[10pt,lettersize,journal,twocolumn,twoside]{IEEEtran}

\usepackage{lettrine}
\usepackage{GoudyIn}
\usepackage{textcomp}
\usepackage{graphicx,amsmath,amsfonts,amsthm,color,comment}
\usepackage{enumitem}
\usepackage{algorithmic}
\usepackage{pgfplots}
\usepackage{pgf}
\usepackage{tikz}
\usepackage{amssymb}
\usepackage{csquotes}
\usepackage{bbm}
\usepackage{multirow}
\usepackage{bm}
\usepackage{cite}
\usepackage{placeins}
\usepackage{dblfloatfix}

\usepackage[nolist]{acronym}
\theoremstyle{definition}

\usepackage{siunitx}

\usepackage[ruled,linesnumbered,vlined]{algorithm2e}
\SetNlSty{bfseries}{\color{black}}{}
\SetArgSty{textnormal}
\pgfplotsset{compat=1.15}
\SetKw{Continue}{continue}
\SetKw{Break}{break}

\def\ben{\begin{eqnarray*}}
\def\een{\end{eqnarray*}}

\def\H5G{H_\text{CA-Polar}}

\begin{document}

\title{RLNC vs Reed-Solomon in Satellite Channels: Crossover Analysis and the Bridging Property}

\author{Angeline Szczepanski, \IEEEauthorblockN{Vipindev Adat Vasudevan  
}
\thanks{Angeline Szczepanski (aszczepanski2010@gmail.com) is with the International School of Boston, Cambridge, MA, USA and the Global Young Researchers' Academy (GYRA), Oregon, USA. Vipindev Adat Vasudevan (vipindevadat@gmail.com) is with the Global Young Researchers' Academy (GYRA), Oregon, USA. \\ This manuscript is currently a preprint under journal review.}
}

\maketitle

\maketitle

\begin{abstract}
Satellite communication systems are increasingly relied upon for connectivity in remote areas and are expected to play a pivotal role in 6G networks, operating under high round-trip times and packet loss where forward error correction is essential. This work evaluates Reed-Solomon (RS) coding, Random Linear Network Coding (RLNC), and a combined RS+RLNC hybrid over a three-hop satellite channel whose loss statistics are derived from orbital geometry and link-budget calculations rather than assumed, comparing RS and RLNC at matched code rates of 0.800. The central finding is the \emph{bridging property} of RLNC: every RS code rate collapses above a crossover threshold $p^*$, while RLNC operates continuously through all failure regions with overhead growing as $1/(1-p)$. At rate~0.800, $p^* \approx 0.21$. Under the derived satcom channel with mean burst duration 166~minutes, RS$(255,204)$ is impractical despite an average loss rate of 1\%, while RLNC requires 15.5~Mbit to recover all 1000 packets. An RS+RLNC hybrid reduces overhead by 19\% relative to pure RLNC at BER$=10^{-3}$. To the authors' knowledge, this is the first analytical derivation of the rate-normalised crossover $p^*$ between RS and RLNC and the first experimental confirmation of the bridging property under a physically derived satellite channel model.
\end{abstract}

\begin{IEEEkeywords}
Random Linear Network Coding, Reed-Solomon coding, Satellite communications, Gilbert-Elliott channel, Forward error correction.
\end{IEEEkeywords}

\section{Introduction}

\lettrine[lines=2]{\textbf{S}}{atellite} communication systems are increasingly relied upon to provide connectivity in remote areas and are expected to play a pivotal role in next-generation non-terrestrial networks \cite{kodheli2021satcom, giordani2021ntn}. Reliable data delivery in such systems presents a fundamental challenge: the channel is inherently hostile, time-varying, and prone to burst losses driven by orbital geometry. As the elevation angle of a satellite relative to a ground station varies along its orbital arc, the link transitions between high-margin availability and complete outage, producing structured burst erasure patterns that are poorly captured by memoryless channel models \cite{gilbert1960capacity, fontan2022review}. The time-varying nature of the link is governed by a physical chain connecting orbital position to link margin and ultimately to packet loss probability, a chain rooted in free-space path loss, antenna geometry, and receiver noise temperature \cite{geisel1990tutorial}. Designing efficient error correction schemes for such channels requires both a realistic physical model and a rigorous analytical framework for comparing competing coding strategies.

In such channels, retransmission-based reliability mechanisms such as ARQ and HARQ are limited by high round-trip times (RTTs). Prior work has shown that network coding can outperform retransmission-based reliability in bursty wireless and satellite environments, motivating packet-level forward erasure correction schemes that recover losses without feedback \cite{landon2025rethinking, vasudevan2026revisiting}.

Two families of forward error correction codes are of particular relevance to satellite communications. Reed-Solomon codes are classical algebraic block codes that operate at the symbol level: they encode $K$ information symbols into a codeword of $N$ symbols and can correct up to $\rho = N-K$ erasures within the received codeword \cite{reed1960polynomial}. They have been adopted as the standard by the CCSDS and used in satellite broadcast standards including DVB \cite{ccsds2017, etsi1997dvb}, and have a long history of deployment in deep space and satellite links \cite{sklar2001reedsolomon, morello2006dvb}. Random Linear Network Coding (RLNC) is a more recent approach rooted in the algebraic theory of network information flow \cite{ahlswede2000network, koetter2003algebraic} in which coded packets are formed as random linear combinations of original data packets over a Galois field \cite{ho2006random, chou2003practical}. This is a fundamentally different protection mechanism from RS: RS is commonly used as a forward error correction scheme correcting symbol-level errors within a received codeword, while RLNC is a forward erasure correction scheme recovering entirely lost packets at the network layer. RLNC belongs to the broader family of rateless and fountain codes \cite{luby2002lt, shokrollahi2006raptor}, which share the property that the encoder need not know the loss rate in advance and the receiver decodes once sufficient coded symbols are accumulated. RLNC has demonstrated throughput and latency advantages in satellite, multicast, and high-RTT network settings \cite{saxena2013network, cocco2015network, lun2006coding, speidel2015network}.

Despite the practical relevance of both schemes, prior comparisons have not rigorously characterised the crossover between RS and RLNC under physically grounded satellite channel conditions. Existing studies often compare the schemes at different code rates, conflating structural advantages with rate effects. RLNC performance under burst channels has been studied using Gilbert-Elliott models \cite{biyikoglu2025modeling}, which are also standard for satellite burst-loss characterisation \cite{fontan2022review, mushkin1989capacity, elliott1963estimates}. However, the crossover loss rate $p^*$ between RS and RLNC at matched rates, and the reason RLNC remains practical beyond every RS failure region, have not been analytically characterised.

This paper addresses that gap. The central novelty is a result we believe has not been established in prior literature: a rigorous analytical characterisation of the crossover loss rate $p^*$ at which RLNC becomes more efficient than RS at the \emph{same code rate}, and the proof that no RS code rate can eliminate this crossover. Prior comparisons of RS and RLNC have either used different code rates \cite{saxena2013network}, conflating structural differences with rate effects, compared achievable rate rather than transmission overhead \cite{saxena2015llsnc}, or have not derived $p^*$ analytically. The specific contributions are: \begin{itemize} \item \textbf{Crossover analysis:} We derive the crossover loss rate $p^*$ between RS and RLNC at matched rate 0.800, obtaining $p^* \approx 0.21$. \item \textbf{Bridging property:} We prove analytically and confirm experimentally that every RS code collapses above its own finite $p^*$ while RLNC operates continuously through all such failure regions with a single fixed scheme, spanning the failure regions of every RS code rate evaluated in this work. \item \textbf{Burst penalty quantification:} We show that under the physically derived satcom channel with mean burst duration 166~minutes, RS$(255,204)$ is impractical despite $\bar{\epsilon} = 1\%$, while RLNC requires only 15.5~Mbit to recover all 1000 packets, confirming that $\bar{\epsilon}$ alone is a misleading metric for code selection on satellite channels. \item \textbf{Hybrid scheme evaluation:} A concatenated RS+RLNC hybrid reduces overhead by approximately 19\% relative to pure RLNC at BER$=10^{-3}$, providing simultaneous bit error and burst erasure correction at the cost of approximately $2\times$ overhead on the pure-erasure satcom channel.
\end{itemize}

Beyond CCSDS, RS(255,239) also underlies the DVB broadcast standard, deployed in shortened form as RS(204,188)~\cite{etsi300744}, confirming that the rates evaluated in this work span real deployed systems rather than arbitrary choices. Recent work continues to study the RS rate/protection tradeoff directly: \cite{wondimu2025performance} shows a higher-rate, less protected RS code outperforming a lower-rate, more-protected one in a concatenated coding context, the same tension this paper's crossover analysis formalizes. Older satellite specific studies confirm RS coding gain scales sharply with channel severity rather than average loss alone~\cite{attia1999fcma}, directly supporting the burst-penalty argument in Section~\ref{sec:burst_penalty}.

The remainder of this paper is organised as follows. Section~\ref{sec:system} presents the satellite scenario and derives the time-varying channel model. Section~\ref{sec:fec} describes the coding schemes. Section~\ref{sec:analysis} develops the mathematical performance framework including the crossover analysis and bridging property. Section~\ref{sec:simulation} presents simulation results. Section~\ref{sec:discussion} discusses implications and limitations. Section~\ref{sec:conclusion} concludes.

\section{System Model and Channel Characterization}
\label{sec:system}

\subsection{Satellite Scenario}

The following scenario is implemented as a simulation using the MATLAB Satellite Communications Toolbox. All results in this paper are simulation-based.

The satellite communication system considered in this work consists of two satellites and two ground stations, forming a three-hop relay path. The uplink connects Ground Station~1 (GS1) to Satellite~1 (Sat1), the crosslink connects Sat1 to Satellite~2 (Sat2), and the downlink connects Sat2 to Ground Station~2 (GS2). GS1 is located in Bangalore, India (12.94\textdegree N, 77.69\textdegree E) and GS2 is located in Sydney, Australia (33.80\textdegree S, 151.18\textdegree E). Both satellites follow circular orbits with a semi-major axis of 10,000~km, propagated using a two-body Keplerian model over a 24-hour simulation window with a 60-second sample interval, giving 1440 time steps in total. A minimum elevation mask of 5\textdegree\ is applied at both ground stations: the end-to-end link is considered available only when both GS1--Sat1 and Sat2--GS2 elevation angles exceed this threshold simultaneously.

\subsection{Link Budget Model}

At each time step $k$, the range of each hop is determined by the instantaneous satellite positions in the Earth-Centered Earth-Fixed (ECEF) frame. The peak antenna gain for a circular aperture of diameter $D$ at frequency $f$ is

\begin{equation}
G = 10 \log_{10} \left( \eta \left( \frac{\pi D f}{c} \right)^2 \right)
\label{eq:antenna_gain}
\end{equation}

where $\eta = 0.55$ is the aperture efficiency and $c$ is the speed of light. The free-space path loss for hop $i$ at time step $k$ is

\begin{equation}
\text{FSPL}_i(k) = 20 \log_{10} \left( \frac{4 \pi d_i(k) f_i}{c} \right)
\label{eq:fspl}
\end{equation}

where $d_i(k)$ is the range and $f_i$ is the carrier frequency of hop $i$. The received $E_b/N_0$ at each hop receiver is

\begin{equation}
\left. \frac{E_b}{N_0} \right|_i = P_{T,i} + G_{T,i} - \text{FSPL}_i + \left. \frac{G}{T} \right|_{\text{rx}} + 10 \log_{10} \left( \frac{1}{k_B R_b} \right)
\label{eq:ebno_hop}
\end{equation}

where $P_{T,i}$ is the transmit power (dBW), $G_{T,i}$ is the transmit antenna gain, $G/T$ is the receive figure of merit (dB/K), $k_B$ is Boltzmann's constant, and $R_b = 10^6$~bps is the bit rate. The system assumes regenerative repeaters at each satellite, so the three hops combine in the linear domain as

\begin{equation}
\left. \frac{E_b}{N_0} \right|_{\text{total}} = \left[ \left( \frac{E_b}{N_0} \right)_1^{-1} + \left( \frac{E_b}{N_0} \right)_2^{-1} + \left( \frac{E_b}{N_0} \right)_3^{-1} \right]^{-1}
\label{eq:ebno_combined}
\end{equation}

The weakest hop dominates this combination. The end-to-end link margin is

\begin{equation}
\Lambda(k) = \left. \frac{E_b}{N_0} \right|_{\text{total}} - \left. \frac{E_b}{N_0} \right|_{\text{req}}
\label{eq:link_margin}
\end{equation}

where the required $E_b/N_0 = 1$~dB. Key link budget parameters are summarised in Table~\ref{tab:link_budget}.

\begin{table}[t]
\centering
\caption{Link Budget Parameters}
\label{tab:link_budget}
\begin{tabular}{lll}
\hline
\textbf{Parameter} & \textbf{Value} & \textbf{Description} \\
\hline
$f_{\text{uplink}}$      & 30 GHz  & GS1 to Sat1 frequency \\
$f_{\text{crosslink}}$   & 30 GHz  & Sat1 to Sat2 frequency \\
$f_{\text{downlink}}$    & 27 GHz  & Sat2 to GS2 frequency \\
$P_{T,\text{GS}}$        & 30 dBW  & Ground station transmit power \\
$P_{T,\text{sat}}$       & 15 dBW  & Satellite transmit power \\
$D_{\text{GS}}$          & 2.0 m   & Ground station dish diameter \\
$D_{\text{sat}}$         & 0.5 m   & Satellite dish diameter \\
$\eta$                   & 0.55    & Aperture efficiency \\
$G/T$                    & 3 dB/K  & Receive figure of merit \\
$(E_b/N_0)_{\text{req}}$ & 1 dB    & Required $E_b/N_0$ at GS2 \\
$R_b$                    & 1 Mbps  & Bit rate \\
\hline
\end{tabular}
\end{table}

\subsection{Packet Loss Probability Model}

The link margin $\Lambda(k)$ is mapped to a per-step packet loss probability via a sigmoid function

\begin{equation}
P_{\text{loss}}(k) = \begin{cases} 1 & \text{if elevation} < 5\textdegree \\ P_{\min} + \dfrac{P_{\max} - P_{\min}}{1 + e^{\alpha \Lambda(k)}} &
\text{otherwise}
\end{cases}
\label{eq:sigmoid}
\end{equation}

with $P_{\max} = 0.95$, $P_{\min} = 0.01$, and $\alpha = 0.8$, chosen to reproduce the qualitative shape of empirical BER/margin curves for satellite links \cite{geisel1990tutorial} rather than fitted to measured data. This mapping captures the physical behaviour of a satellite link:

\subsection{Gilbert-Elliott Channel Model}

The time-varying loss sequence $\{P_{\text{loss}}(k)\}$ has an inherent burst structure driven by orbital geometry: outage windows during which $P_{\text{loss}} = 1$ persist for extended periods, while open-link windows with low loss rates last for tens of minutes. This burst structure is not captured by a memoryless i.i.d.\ model and motivates the use of the Gilbert-Elliott (GE) two-state Markov channel model \cite{gilbert1960capacity, elliott1963estimates, mushkin1989capacity}.

The GE model defines a good state $G$ (open link, low loss) and a bad state $B$ (outage or severe degradation, high loss), with transition matrix

\begin{equation}
\mathbf{T} = \begin{pmatrix} 1-p & p \\ r & 1-r \end{pmatrix}
\label{eq:ge_transition}
\end{equation}

where $p = P(G \rightarrow B)$ and $r = P(B \rightarrow G)$. The steady-state probabilities are $\pi_G = r/(p+r)$ and $\pi_B = p/(p+r)$, with mean burst duration $1/r$ and mean good period $1/p$. The marginal packet loss probability is

\begin{equation}
\bar{\epsilon} = \pi_G \epsilon_G + \pi_B \epsilon_B
\label{eq:epsilon_bar}
\end{equation}

where $\epsilon_G$ is the mean loss during open-link windows and $\epsilon_B = 1$ during outage. The GE parameters are estimated from the simulated link availability sequence via maximum likelihood \cite{mushkin1989capacity},

\begin{equation}
\hat{p} = \frac{N_{G \rightarrow B}}{N_G}, \qquad \hat{r} = \frac{N_{B \rightarrow G}}{N_B}
\label{eq:ge_estimation}
\end{equation}

A critical observation is that $\bar{\epsilon}$ alone is insufficient to characterize the channel for block codes. Two channels with identical $\bar{\epsilon}$ but different burst structures perform very differently: rare long bursts are far more damaging to block codes than frequent short bursts at the same average loss rate \cite{biyikoglu2025modeling}. This distinction is central to the crossover analysis in Section~\ref{sec:analysis}.

\section{Forward Error Correction Schemes}
\label{sec:fec}

\subsection{Reed-Solomon Coding}

Reed-Solomon codes are a family of MDS algebraic block codes defined over $\text{GF}(2^m)$ \cite{reed1960polynomial}. With $m = 8$, each symbol is one byte and the maximum codeword length is $N_{\max} = 255$ symbols. A general RS$(N, K)$ code encodes $K$ information symbols into $N$ symbols via the generator polynomial

\begin{equation}
g(x) = \prod_{i=0}^{2t-1} (x - \alpha^i)
\label{eq:rs_generator}
\end{equation}

where $\alpha$ is a primitive element of $\text{GF}(2^8)$. RS codes achieve the Singleton bound $d_{\min} = N - K + 1$ \cite{reed1960polynomial}, giving error and erasure correction capacities

\begin{equation}
t = \left\lfloor \frac{N - K}{2} \right\rfloor, \qquad \rho = N - K
\label{eq:rs_capacity}
\end{equation}

RS codes are forward error correction (FEC) codes in the strict sense: they correct symbol-level errors within a received codeword. On a pure erasure channel, an RS block decodes if and only if at most $\rho$ symbols are lost. The code rate $R_{\text{RS}} = K/N$ is fixed at construction time and cannot be changed without rebuilding the code entirely \cite{sklar2001reedsolomon}. This rigidity is a key structural difference from RLNC. In this work we evaluate three RS configurations using $N = 255$ throughout, chosen so that the comparison is not confounded by different block lengths: RS$(255,223)$ with $\rho=32$ and rate $0.875$ (the CCSDS standard \cite{ccsds2017}), RS$(255,204)$ with $\rho=51$ and rate $0.800$ (exactly matched to RLNC), and RS$(255,191)$ with $\rho=64$ and rate $0.749$. Table~\ref{tab:rs_codes} lists these alongside additional standard configurations for reference.

\begin{table}[t]
\centering
\caption{Reed-Solomon Code Configurations Evaluated}
\label{tab:rs_codes}
\begin{tabular}{lllllll}
\hline
\textbf{Code} & \textbf{Field} & $N$ & $K$ & $t$ & $\rho$ &
\textbf{Rate} \\
\hline
RS(7,3)     & GF($2^3$) & 7   & 3   & 2  & 4  & 0.429 \\
RS(255,191) & GF($2^8$) & 255 & 191 & 32 & 64 & 0.749 \\
RS(255,204) & GF($2^8$) & 255 & 204 & 25 & 51 & 0.800 \\
RS(255,223) & GF($2^8$) & 255 & 223 & 16 & 32 & 0.875 \\
RS(255,239) & GF($2^8$) & 255 & 239 & 8  & 16 & 0.937 \\
\hline
\end{tabular}
\end{table}

\subsection{Random Linear Network Coding}\label{sec:rlnc_innovative}

RLNC is a packet-level forward erasure correction scheme rooted in the algebraic theory of network coding \cite{koetter2003algebraic, ahlswede2000network}. The encoder groups original packets into generations and transmits random linear combinations over $\text{GF}(2^8)$ \cite{ho2006random, chou2003practical}. This distinguishes it fundamentally from RS: while RS corrects symbol errors within a received block, RLNC recovers entirely lost packets by accumulating linearly independent coded packets across multiple rounds. RLNC belongs to the broader family of rateless codes \cite{luby2002lt, shokrollahi2006raptor}, which share the property that the encoder need not be configured for a specific loss rate and the receiver decodes once sufficient coded packets are accumulated. Each transmitted coded packet is

\begin{equation}
\mathbf{c} = \sum_{i=1}^{K_{\text{gen}}} \gamma_i \mathbf{u}_i, \qquad \gamma_i \overset{\$}{\leftarrow} \text{GF}(2^8)
\label{eq:rlnc_encoding}
\end{equation}

Decoding succeeds when the receiver accumulates sufficient packets to satisfy $\text{rank}(\mathbf{C}) = K_{\text{gen}}$, at which point Gaussian elimination over $\text{GF}(2^8)$ recovers all original packets. Each received packet is innovative with probability

\begin{equation}
P(\text{innovative} \mid \text{rank} = r) = 1 - 2^{-8(K_{\text{gen}} - r)}
\label{eq:rlnc_innovative}
\end{equation}

For $K_{\text{gen}} = 16$ over $\text{GF}(2^8)$, this probability exceeds $1 - 2^{-8} \approx 0.996$ at every rank, making linear dependence failures negligible in practice \cite{ho2006random, lun2006coding, heide2020rlnc}.

Block RLNC is chosen in this work because satellite links have large RTTs, making per-packet feedback inefficient \cite{landon2025rethinking, vasudevan2026revisiting, esmaeilzadeh2014joint}. With block RLNC, the encoder transmits a full generation of $N_{\text{coded}} = 20$ coded packets per round without waiting for feedback. A burst that destroys one round contributes only a delay, not a failure: the encoder simply sends an additional round of $N_{\text{coded}}$ packets until $K_{\text{gen}}$ is reached. In this work RLNC uses $K_{\text{gen}} = 16$, $N_{\text{coded}} = 20$, giving a \emph{per-round} rate $R_{\text{RLNC}} = 0.800$, matching RS$(255,204)$; unlike RS's fixed block rate, this is not a bound on total overhead, since extra rounds absorb increasing loss, which is what enables the bridging property.

\subsection{RS+RLNC Hybrid Coding}

The RS+RLNC hybrid uses RS as an inner code correcting symbol errors within each received packet, and RLNC as an outer code correcting packet erasures across the channel \cite{vasudevan2026revisiting}. This concatenated scheme provides simultaneous protection against both bit errors and burst erasures, a capability available in neither pure RS nor pure RLNC alone. The only widely deployed alternative combining error and erasure correction is HARQ using LDPC \cite{landon2025rethinking}; the RS+RLNC hybrid provides a structurally different alternative that inherits RLNC's burst resilience at the outer layer. Adaptive inner code selection strategies for such concatenated schemes have been studied in the context of unequal error protection \cite{vukobratovic2012unequal}, motivating the multi-configuration evaluation in Experiment~6. The total code rate is

\begin{equation}
R_{\text{hybrid}} = \frac{K_{\text{in}}}{N_{\text{in}}} \cdot \frac{K_{\text{gen}}}{N_{\text{coded}}}
\label{eq:hybrid_rate}
\end{equation}

On a pure erasure channel, inner RS adds overhead without correction benefit and pure RLNC is optimal. The hybrid becomes advantageous when residual bit errors are present, as characterised in Experiment~6.

\section{Mathematical Performance Analysis}
\label{sec:analysis}

\subsection{RS Decoding Success Probability}
\label{sec:rs_prob}

On the packet erasure channel, each of the $N$ RS packets is independently lost with probability $p$. Decoding succeeds if and only if at most $\rho = N - K$ packets are erased,

\begin{equation}
P_{\text{decode}}^{\text{RS}}(p) = \sum_{j=0}^{\rho} \binom{N}{j} p^j (1-p)^{N-j}
\label{eq:rs_decode_prob}
\end{equation}

For large $N$ and $\rho$, this binomial sum is computed in log-space using the Stirling approximation to avoid numerical overflow \cite{sklar2001reedsolomon}. The transmission overhead ratio, defined as expected transmitted bits per information bit delivered, is

\begin{equation}
\eta_{\text{RS}}(p) = \frac{N/K}{P_{\text{decode}}^{\text{RS}}(p)}
\label{eq:rs_overhead}
\end{equation}

As $p$ increases, $P_{\text{decode}}^{\text{RS}}(p)$ falls rapidly and $\eta_{\text{RS}}(p)$ diverges. The rate of divergence depends strongly on $\rho$: codes with larger $\rho$ tolerate higher loss before collapsing, but every RS code eventually collapses regardless of $\rho$.

\subsection{RLNC Decoding Success Probability}
\label{sec:rlnc_prob}

If $n$ coded packets are received from a generation of size $K_{\text{gen}}$, the probability that they span the full space over $\text{GF}(2^8)$ is \cite{ho2006random, lun2006coding}

\begin{equation}
P_{\text{decode}}^{\text{RLNC}}(n, K) = \prod_{r=0}^{K-1} \left(1 - 2^{8(r-n)}\right), \qquad n \geq K
\label{eq:rlnc_decode_prob}
\end{equation}

For $n = K_{\text{gen}} = 16$, this product is dominated by its smallest factor and evaluates to approximately $1 - 2^{-8} \approx 0.996$, consistent with Section~\ref{sec:rlnc_innovative}, with the linear dependence correction negligible in practice.

\begin{equation}
\mathbb{E}[T_{\text{RLNC}}] \approx \frac{K_{\text{gen}}}{1-p} = \frac{16}{1-p}
\label{eq:rlnc_expected_tx}
\end{equation}

giving overhead ratio

\begin{equation}
\eta_{\text{RLNC}}(p) \approx \frac{N_{\text{coded}}}{K_{\text{gen}}} \cdot \frac{1}{1-p} = \frac{1.25}{1-p}
\label{eq:rlnc_overhead}
\end{equation}

This grows only linearly with $p$. The structural reason is that RLNC accumulates degrees of freedom across multiple rounds: a lost round delays decoding but does not fail the generation, so the effective overhead per delivered packet is governed only by the expected number of rounds needed to accumulate $K_{\text{gen}}$ received packets \cite{lun2006coding}. Equation~\eqref{eq:rlnc_overhead} excludes the per-packet coefficient-vector header, a standard simplification \cite{lun2006coding} that would modestly narrow but not eliminate RLNC's overhead advantage.

\subsection{The Crossover Loss Rate $p^*$ and the Bridging Property}
\label{sec:crossover}

This subsection presents the central analytical contribution of the paper. At low loss rates, RS is more bandwidth-efficient than RLNC when both have similar code rates because RS has a deterministic block structure that requires no per-generation overhead. As loss increases, however, $\eta_{\text{RS}}(p)$ grows without bound while $\eta_{\text{RLNC}}(p)$ grows only as $1.25/(1-p)$. The crossover loss rate $p^*$ is the packet loss probability at which both schemes require equal transmission overhead:

\begin{equation}
\eta_{\text{RS}}(p^*) = \eta_{\text{RLNC}}(p^*)
\label{eq:crossover_def}
\end{equation}

Substituting~\eqref{eq:rs_overhead} and~\eqref{eq:rlnc_overhead},

\begin{equation}
\frac{N/K}{\displaystyle\sum_{j=0}^{\rho}\binom{N}{j}(p^*)^j(1-p^*)^{N-j}} = \frac{N_{\text{coded}}/K_{\text{gen}}}{1-p^*}
\label{eq:crossover_eq}
\end{equation}

This is solved numerically. The crossover $p^*$ depends on the RS code parameters $(N, K, \rho)$ and the RLNC code rate $K_{\text{gen}}/N_{\text{coded}}$. For the rate-matched comparison in this work ($R_{\text{RS}} = R_{\text{RLNC}} = 0.800$), the crossover depends only on $\rho$: higher $\rho$ shifts $p^*$ to a larger value, allowing RS to remain competitive at lower loss rates before RLNC takes over.

\begin{figure}[tp]
\centering
\includegraphics[width=\columnwidth]{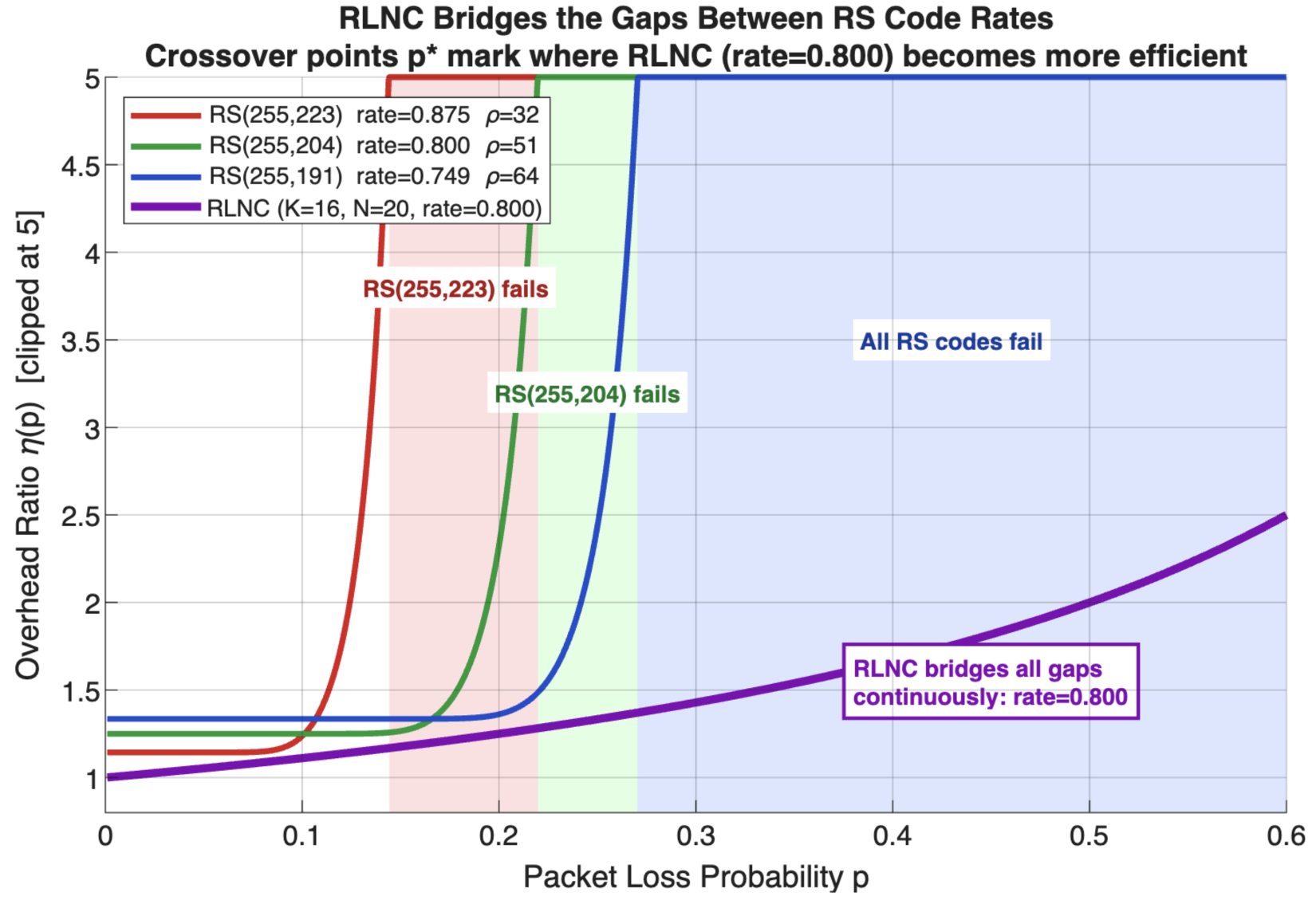}
\caption{The bridging property of RLNC: every RS code rate eventually collapses to infinite overhead at its crossover $p^*$, while RLNC (rate$=0.800$, purple) operates continuously through all failure regions. Shaded regions mark loss ranges where each RS code becomes impractical. At the matched rate of 0.800, RLNC becomes more efficient than RS$(255,204)$ above $p^* \approx 0.21$.}
\label{fig:bridging}
\end{figure}

The key result, which we term the \emph{bridging property} of RLNC, is that $p^*$ is finite for every RS code rate, but there is no corresponding crossover in the other direction for RLNC. Above $p^*$, RLNC is decisively more efficient and the gap grows rapidly. The envelope of the best available RS code rate at each loss level collapses by $p \approx 0.27$ at the matched rate of 0.800, beyond which RLNC is the only practical option regardless of which RS code rate is selected. This property distinguishes RLNC not just as a better-performing scheme at high loss but as a structurally different class of solution that cannot be matched by any fixed block code regardless of its parameters. Fig.~\ref{fig:bridging} illustrates this property.

The numerical solution of~\eqref{eq:crossover_eq} is validated directly against simulation in Section~\ref{sec:exp3}: the closed-form expressions~\eqref{eq:rs_overhead} and~\eqref{eq:rlnc_overhead} predict simulated transmission cost to within 1\% for RS and within 5\% for RLNC at low-to-moderate loss across four RS code rates.

\subsection{Burst Channel Penalty: RS vs.\ RLNC}
\label{sec:burst_penalty}

The crossover analysis above assumes a memoryless i.i.d.\ erasure channel. Under the GE model, losses are correlated and burst length follows a geometric distribution \cite{gilbert1960capacity},

\begin{equation}
L_B \sim \text{Geom}(r), \qquad \mathbb{E}[L_B] = \frac{1}{r}
\label{eq:burst_length}
\end{equation}

For RS$(N,K)$ with erasure correction capacity $\rho$, decoding fails with certainty whenever a burst exceeds $\rho$ consecutive steps,

\begin{equation}
P(\text{RS block fails} \mid \text{burst}) \geq P(L_B > \rho) = (1 - r)^{\rho}
\label{eq:rs_burst_fail}
\end{equation}

This is independent of the per-packet loss probability $p$: even if $p$ is very small, a sufficiently long burst guarantees RS decoding failure. The marginal loss $\bar{\epsilon}$ therefore systematically understates the effective overhead of RS under the GE model \cite{biyikoglu2025modeling}.

RLNC does not share this vulnerability. A burst that destroys one round of $N_{\text{coded}} = 20$ packets contributes only a temporary delay: the receiver continues accumulating innovative packets in subsequent rounds \cite{lun2006coding}. RLNC decoding fails only if the total innovative packets received across all rounds falls below $K_{\text{gen}} = 16$, which requires sustained loss across many consecutive rounds rather than a single burst. This asymmetry is summarised as

\begin{equation}
\eta_{\text{RS}}^{\text{eff}} \gg \eta_{\text{RS}}(\bar{\epsilon}), \qquad \eta_{\text{RLNC}}^{\text{eff}} \approx \eta_{\text{RLNC}}(\epsilon_G)
\label{eq:ge_asymmetry}
\end{equation}

This burst penalty, and the way it compounds the crossover effect to make RS impractical even at low average loss rates, is highlighted as a key contribution in the introduction and validated experimentally in Experiments~4 and~5 \cite{biyikoglu2025modeling, yang2014throughput}.

This is the central analytical result of the paper: the effective overhead of RS under a burst channel is far greater than its overhead on an i.i.d.\ channel at the same average loss rate, while RLNC's effective overhead remains approximately equal to its i.i.d.\ overhead evaluated at the within-window loss rate $\epsilon_G$. The burst penalty compounds the crossover effect: the burst structure of the satellite channel causes RS to effectively operate as if the loss rate were far higher than $\bar{\epsilon}$, pushing it above its crossover threshold even at low average loss rates.

\section{Simulation Setup and Results}
\label{sec:simulation}

\subsection{Simulation Implementation}

A packet-level simulation was implemented in MATLAB using the Satellite Communications Toolbox. Packets are either fully received or fully erased on the erasure channel; bit-level errors are introduced explicitly only in Experiment~6. Performance is measured by the total transmitted bits required to recover 950, 990, and 1000 original packets. The simulation operates in three modes: a fixed packet loss rate sweep over $P \in \{0, 0.01, 0.02, 0.05, 0.1, 0.2, 0.3, 0.4, 0.5, 0.6\}$, a satcom time-varying mode in which $P_{\text{loss}}(k)$ is derived from the orbital geometry of Section~\ref{sec:system}, and a Gilbert-Elliott burst sweep in which $r$ is varied to control mean burst duration. Six experiments are conducted, summarised in Table~\ref{tab:experiments}. All RS configurations use $N = 255$ for a consistent block length comparison. RLNC uses $K_{\text{gen}} = 16$ and $N_{\text{coded}} = 20$ throughout, giving rate $0.800$ exactly matching RS$(255,204)$.

\begin{table}[t]
\centering
\caption{Summary of Simulation Experiments}
\label{tab:experiments}
\begin{tabular}{lll}
\hline
\textbf{Exp.} & \textbf{Channel} & \textbf{Schemes} \\
\hline
1 & Fixed-$p$ sweep     & RS(255,204) rate=0.800 vs theory \\
2 & Fixed-$p$ sweep     & RLNC rate=0.800 vs theory \\
3 & Fixed-$p$ sweep     & Three RS codes vs RLNC, crossover $p^*$ \\
4 & Satcom time-varying & RS(255,204) vs RLNC, burst penalty \\
5 & GE burst sweep      & RS(255,204) vs RLNC, burst duration \\
6 & Fixed-$p$ + satcom  & RS+RLNC hybrid vs pure RLNC \\
\hline
\end{tabular}
\end{table}

The RS simulation transmits full blocks of $N = 255$ packets per attempt and counts a block as successful if and only if the number of lost packets does not exceed $\rho$. A transmission cap of $10^9$ bits is applied. The RLNC simulation uses the fast approximation justified in Section~\ref{sec:rlnc_prob}: received packets are counted as innovative directly, since the linear dependence probability is negligible over $\text{GF}(2^8)$ \cite{ho2006random, lun2006coding}.

\subsection{Experiments 1 and 2: Fixed Packet Loss Rate Results}
\label{sec:fixed}

\begin{figure}[tp]
\centering
\includegraphics[width=\columnwidth]{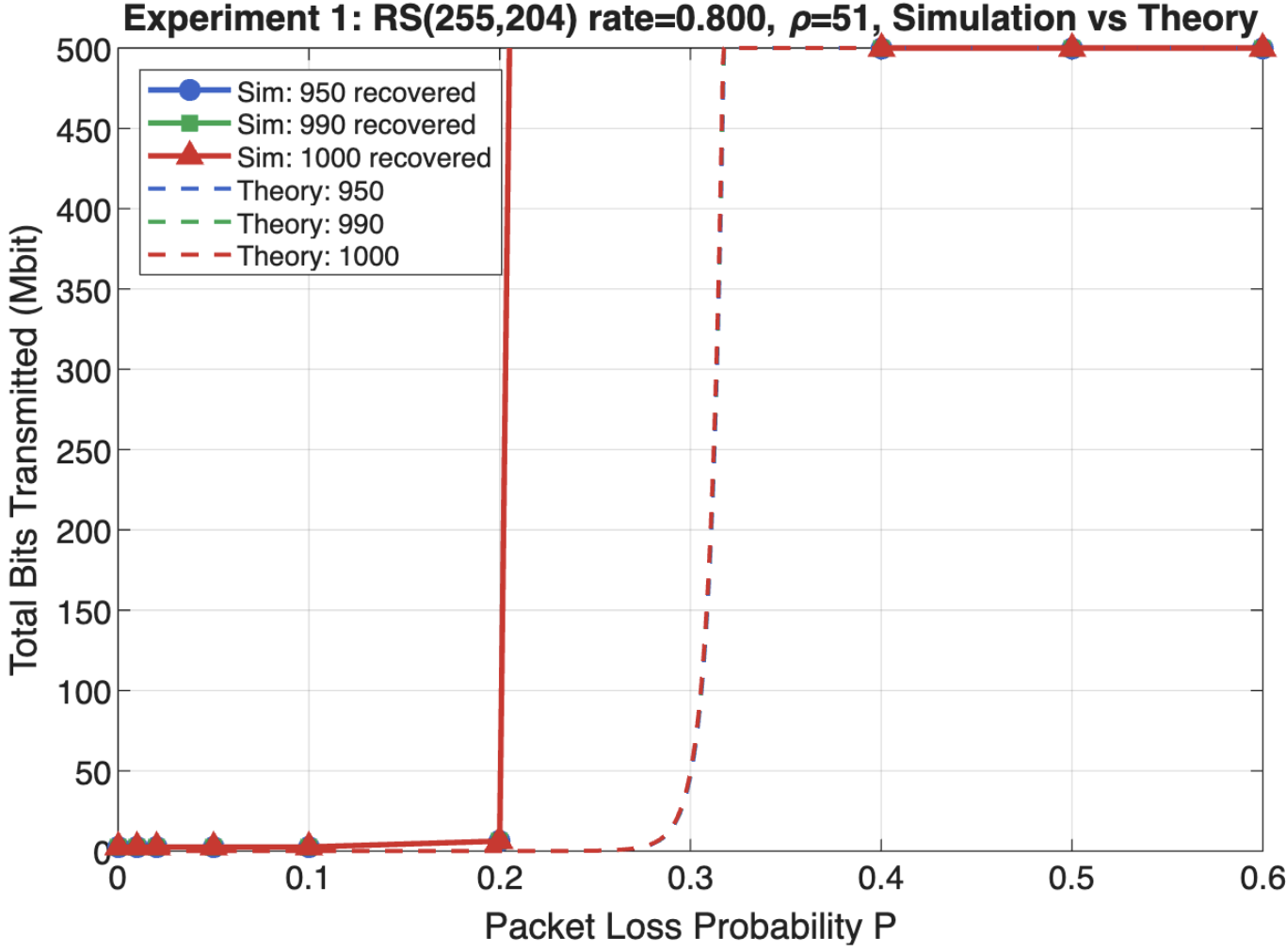}
\caption{Experiment 1: RS$(255,204)$ simulation (solid) vs theory (dashed) at rate$=0.800$, $\rho=51$. RS remains practical to $p \approx 0.20$ before the block failure probability approaches 1 and transmitted bits hit the 500~Mbit ceiling.}
\label{fig:exp1}
\end{figure}

\begin{figure}[tp]
\centering
\includegraphics[width=\columnwidth]{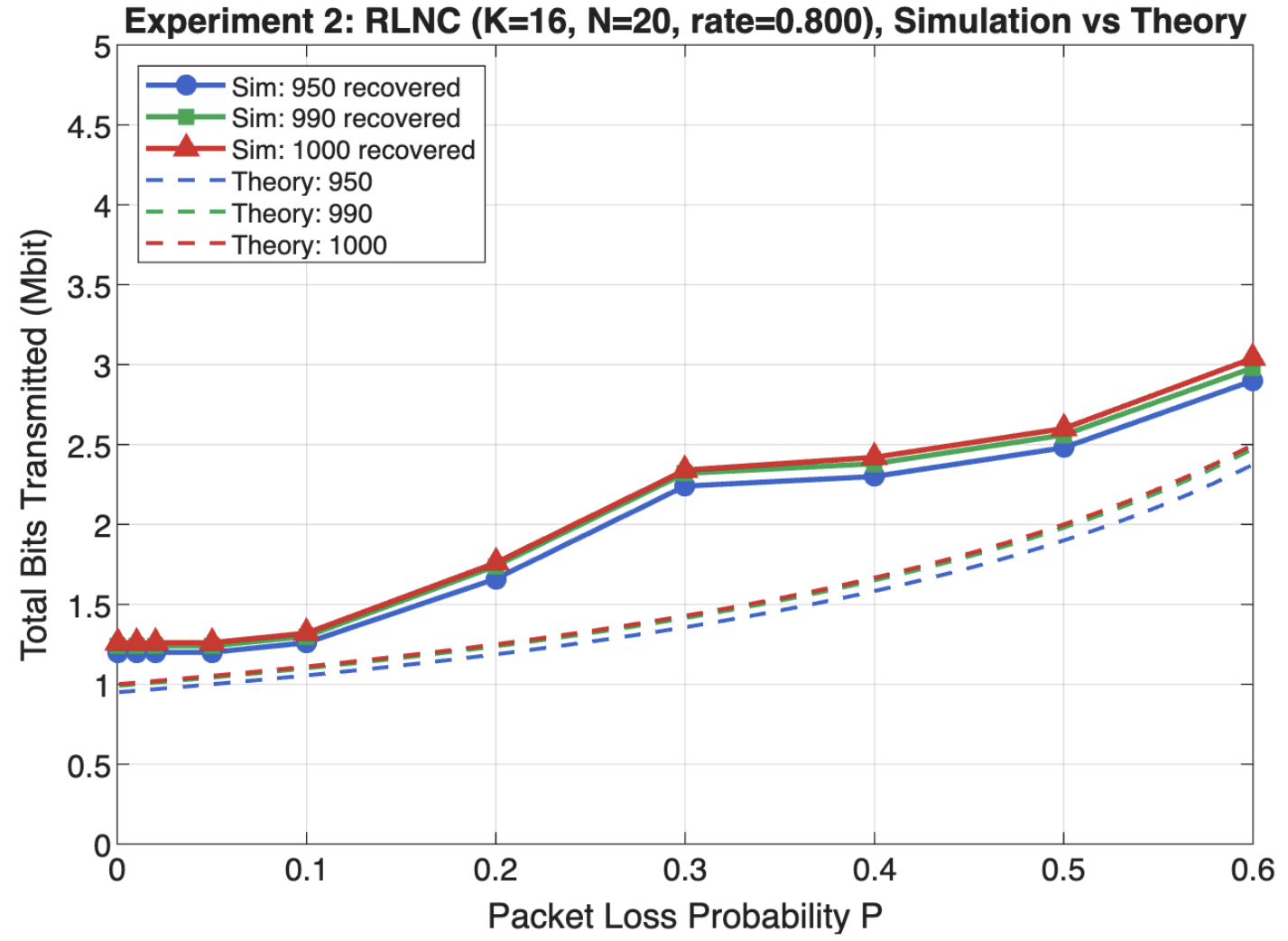}
\caption{Experiment 2: RLNC simulation (solid) vs theoretical prediction $\eta_{\text{RLNC}}(p) \approx 1.25/(1-p)$ (dashed) at rate$=0.800$. The simulation sits slightly above theory due to per-generation start-up overhead for finite generation sizes \cite{ho2006random}.}
\label{fig:exp2}
\end{figure}

Figs.~\ref{fig:exp1} and~\ref{fig:exp2} show the total transmitted bits versus fixed packet loss rate for RS$(255,204)$ and RLNC respectively, both at rate $0.800$.

For RS$(255,204)$, the larger erasure capacity $\rho = 51$ compared to RS$(127,125)$ with $\rho = 2$ means the code remains practical to a much higher loss rate. The collapse occurs at $p \approx 0.20$, where the expected number of losses per block ($255 \times 0.20 = 51$) equals $\rho$ exactly. Above this threshold the block failure probability rises sharply. The simulation closely tracks the theoretical prediction of~\eqref{eq:rs_overhead} before the cap is reached, validating the analytical framework. The three recovery thresholds (950, 990, 1000 packets) are nearly coincident because block-level failure dominates over the specific recovery target.

For RLNC, the simulation sits slightly above the theoretical $1.25/(1-p)$ curve across the full sweep. This gap is expected for finite generation sizes: the theory assumes the receiver can start a new round immediately, but with $K_{\text{gen}} = 16$ packets per generation there is a small additional overhead at generation boundaries \cite{lun2006coding}. The key observation is that RLNC overhead remains bounded and smoothly growing across the entire range $p \in [0, 0.6]$, in sharp contrast to the RS collapse observed in Fig.~\ref{fig:exp1}.

\subsection{Experiment 3: Crossover Points and the Bridging Property}
\label{sec:exp3}

\begin{figure}[tp]
\centering
\includegraphics[width=\columnwidth]{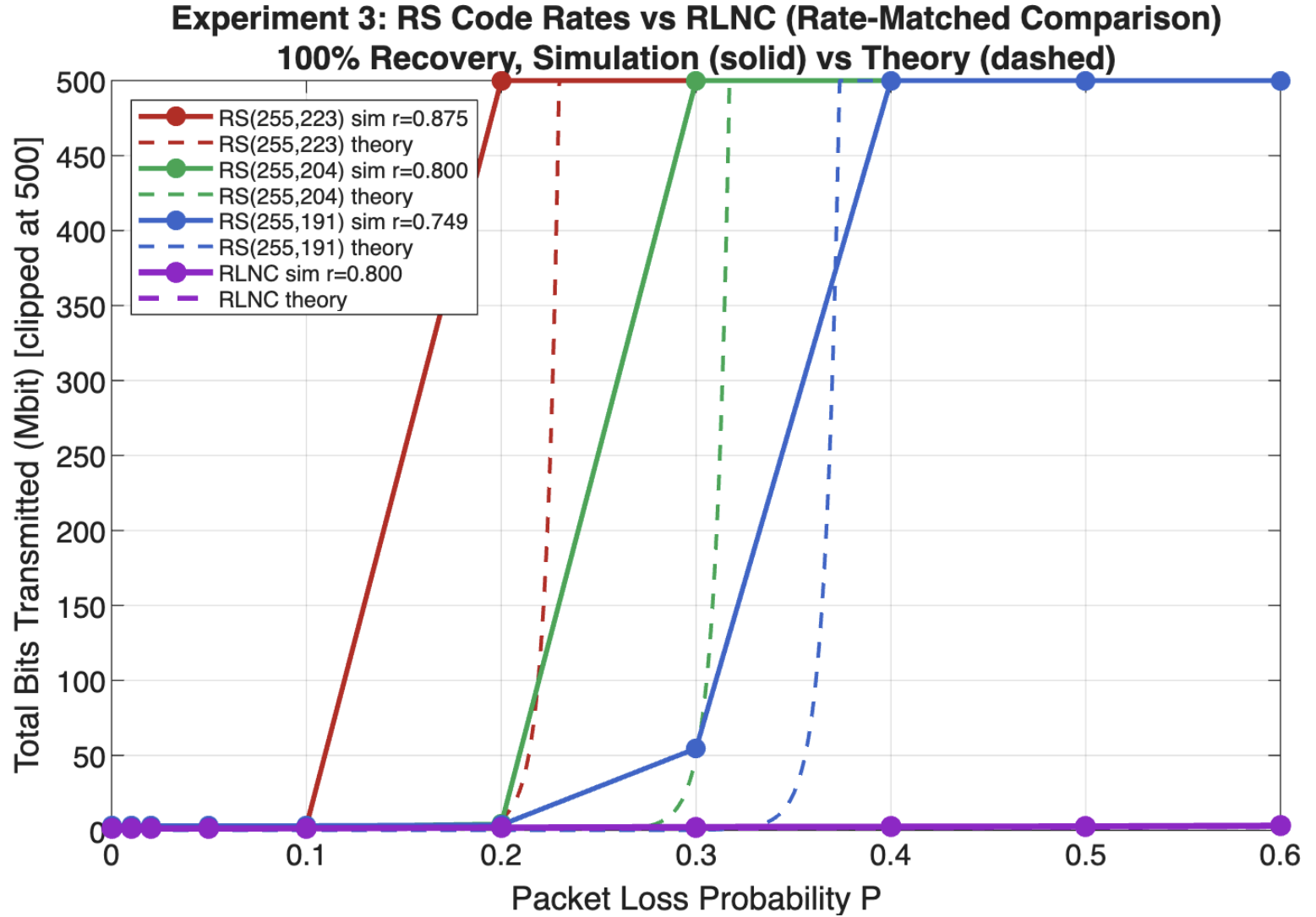}
\caption{Experiment 3: three RS code rates vs RLNC at rate$=0.800$. All RS codes collapse in sequence while RLNC remains below 8~Mbit across the entire range. At the matched rate, RLNC and RS$(255,204)$ begin with nearly identical overhead at low loss, then RLNC becomes more efficient above $p^* \approx 0.21$; $p^*$ shifts right for codes with larger $\rho$, consistent with~\eqref{eq:crossover_eq}.}
\label{fig:exp3_full}
\end{figure}

Experiment 3 is the central experimental result of the paper. It addresses the question: at the same code rate, does RLNC outperform RS, and if so above what crossover loss rate $p^*$? Fig.~\ref{fig:exp3_full} shows the answer.

Fig.~\ref{fig:exp3_full} shows the staircase pattern of RS failures as packet loss increases: RS$(255,223)$ collapses first ($\rho=32$), followed by RS$(255,204)$ and RS$(255,191)$ ($\rho=64$), with larger $\rho$ shifting the collapse threshold rightward, though every RS configuration eventually becomes impractical. At the matched rate of $0.800$, RS$(255,204)$ is slightly more efficient below $p^* \approx 0.21$ due to its deterministic block structure; above this threshold RLNC becomes more efficient as generations continue accumulating across rounds while RS block failures increase rapidly. Since both schemes operate at identical rates, the crossover is structural rather than rate-driven \cite{lun2006coding}; the piecewise-linear appearance between sampled $P$ values reflects sweep granularity, not a discontinuity in the underlying process.

These results confirm the bridging property of RLNC: increasing $\rho$ shifts the RS crossover threshold but does not eliminate it, while a single RLNC configuration operates continuously across all corresponding RS failure regions without reconfiguration.

Table~\ref{tab:theory_vs_sim} quantifies theory-simulation agreement at $p=0.1$ across four RS rates, with RLNC rate-matched to each (unlike the fixed rate 0.800 in Experiments~1--2). RS theory matches simulation within 1\% throughout; RLNC matches within 5\% for rates 0.502--0.800, diverging to 38\% at rate 0.937 where the small generation size ($K_{\text{gen}}=19$, $N_{\text{coded}}=20$) amplifies the finite-generation effect from Section~\ref{sec:fixed}.

\begin{table}[t]
\centering
\caption{Simulated vs.\ Theoretical Transmission Cost (Mbit), $p=0.1$}
\label{tab:theory_vs_sim}
\begin{tabular}{lcccc}
\hline
\textbf{RS Rate} & \textbf{RS sim} & \textbf{RS theory} & \textbf{RLNC sim} & \textbf{RLNC theory} \\
\hline
0.502 & 1.9931 & 1.9922 & 2.0160 & 2.2222 \\
0.702 & 1.4260 & 1.4246 & 1.4519 & 1.5972 \\
0.800 & 1.2505 & 1.2500 & 1.3145 & 1.3889 \\
0.937 & 43.484 & 43.203 & 1.6312 & 1.1806 \\
\hline
\end{tabular}
\end{table}

\subsection{Multiple RLNC Operating Points Within the Bridging Region}
\begin{figure}[tp]
\centering
\includegraphics[width=\columnwidth]{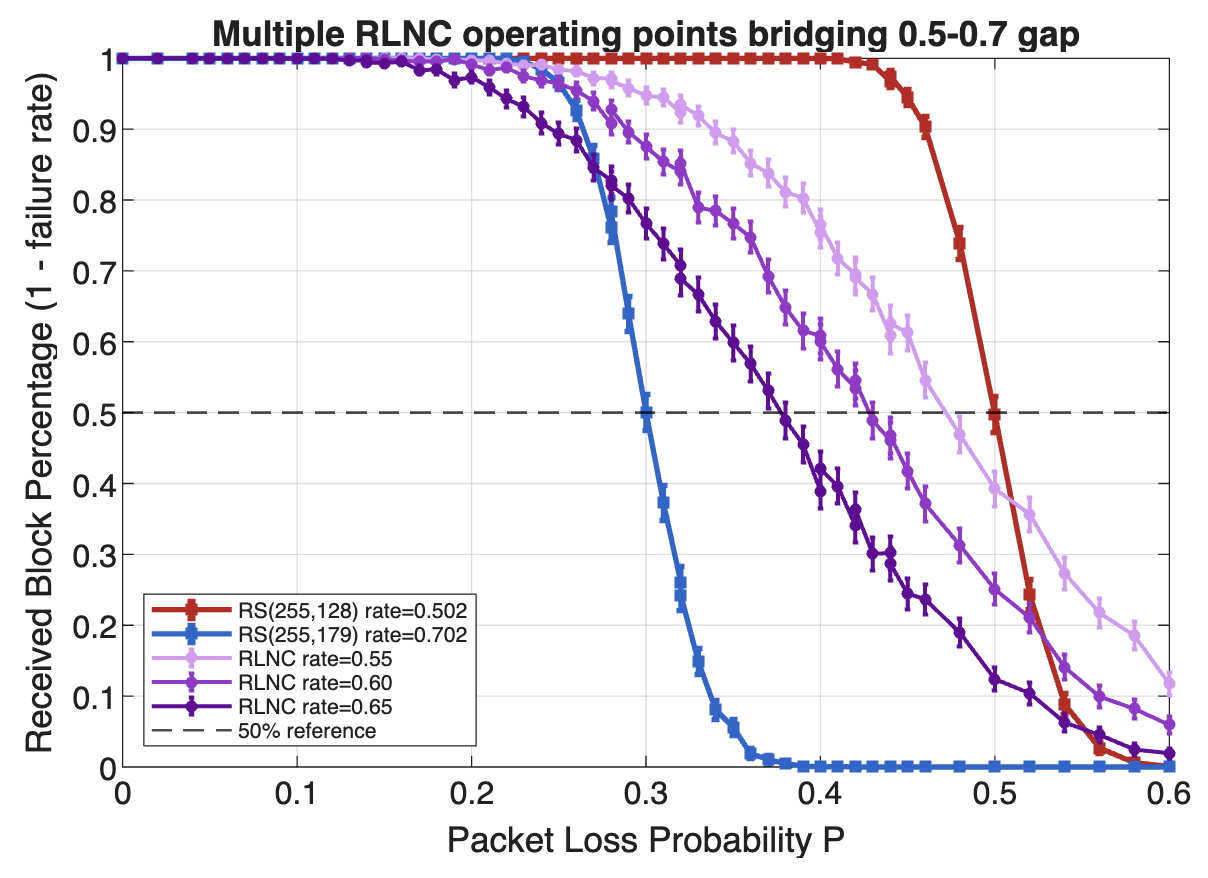}
\caption{Multiple RLNC operating points (rates 0.55, 0.60, 0.65) bridging the gap between RS(255,128) rate=0.502 and RS(255,179) rate=0.702, each crossing 50\% received block percentage at a distinct loss probability.}
\label{fig:multi_rlnc}
\end{figure}

The bridging property in Fig.~\ref{fig:exp3_full} uses a single RLNC rate. Fig.~\ref{fig:multi_rlnc} shows that multiple RLNC rates can be selected within a given RS rate gap, each yielding a distinct crossover point, demonstrating that the bridging region is continuously fillable rather than characterized by one fixed operating point.

\subsection{Experiment 4: Satcom Time-Varying Channel}

The fixed-$p$ sweep of Experiment~3 assumes independent per-packet losses. The satcom channel violates this assumption severely. The GE parameters estimated from the orbital data are $\hat{p} = 0.0734$ and $\hat{r} = 0.0060$, giving a mean burst duration of $1/\hat{r} = 166.5$~minutes. The mean open-link loss rate is $\epsilon_G = 0.0100$. The marginal packet loss rate including outage periods is $\bar{\epsilon} = \pi_G \epsilon_G + \pi_B \epsilon_B$; given $\pi_B = \hat{p}/(\hat{p}+\hat{r}) \approx 0.924$, the true  $\bar{\epsilon} \approx 0.924$, reflecting the fact that the link is in outage for the majority of the 24-hour window. The open-link loss rate $\epsilon_G = 0.010$ is the relevant quantity for assessing RS performance during available windows, and it is this rate at which theory predicts RS should be highly efficient.

\begin{figure}[tp]
\centering
\includegraphics[width=\columnwidth]{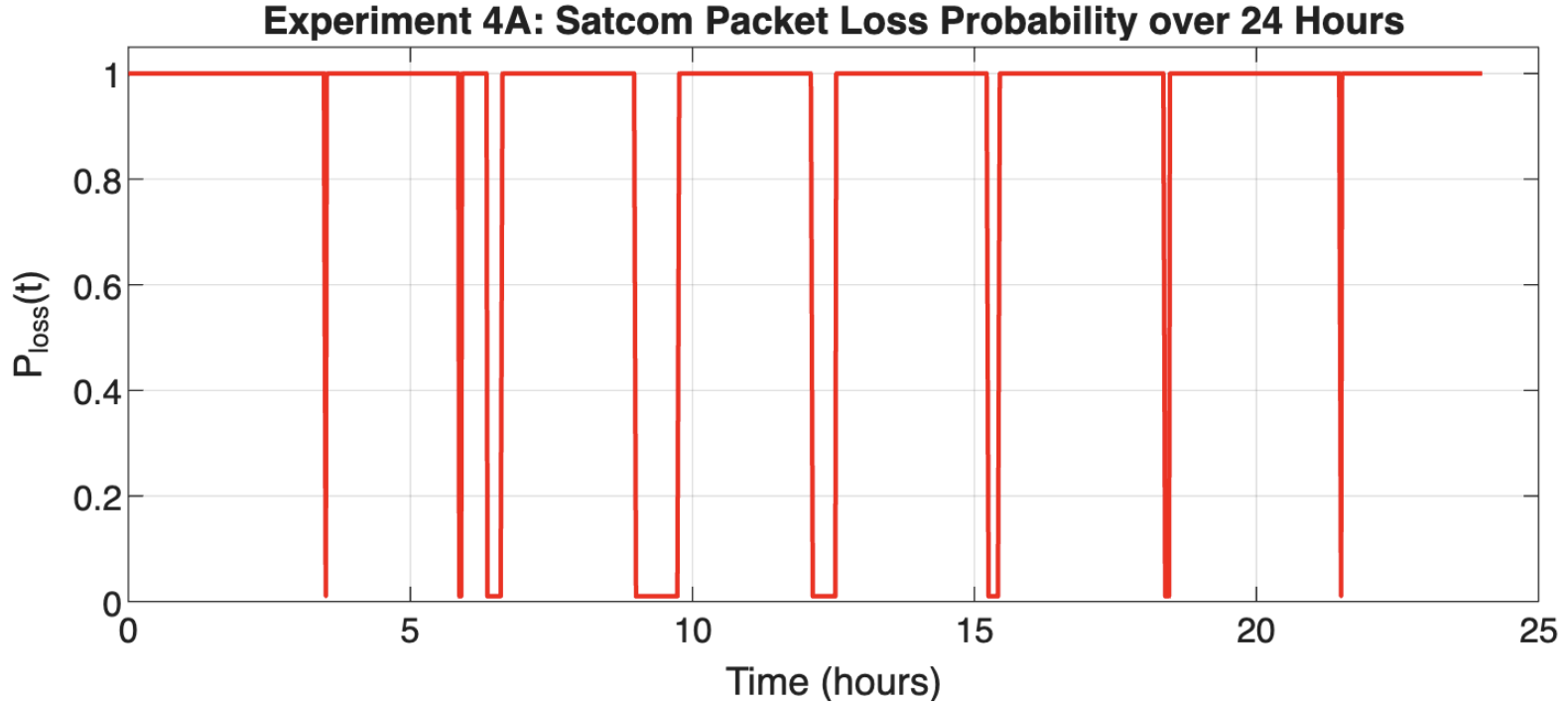}
\caption{Experiment 4A: Satcom packet loss probability over 24~hours, derived from orbital geometry and link budget calculations. $P_{\text{loss}}$ alternates between near-zero during open-link windows and exactly~1 during outage windows. The mean burst duration of 166~minutes far exceeds $\rho = 51$ steps of RS$(255,204)$, guaranteeing RS block failure during every outage window.}
\label{fig:exp4a}
\end{figure}

Fig.~\ref{fig:exp4a} shows the time-varying structure of the channel. The pattern of outage and open-link periods is clearly visible. Each outage window lasts on the order of 166~minutes on average, corresponding to approximately 166 time steps. For RS$(255,204)$ with $\rho = 51$, any burst longer than 51 steps guarantees block failure. Since the mean burst duration of 166 steps is more than three times $\rho$, essentially every outage window causes complete RS decoding failure regardless of average loss rate.

\begin{figure}[tp]
\centering
\includegraphics[width=\columnwidth]{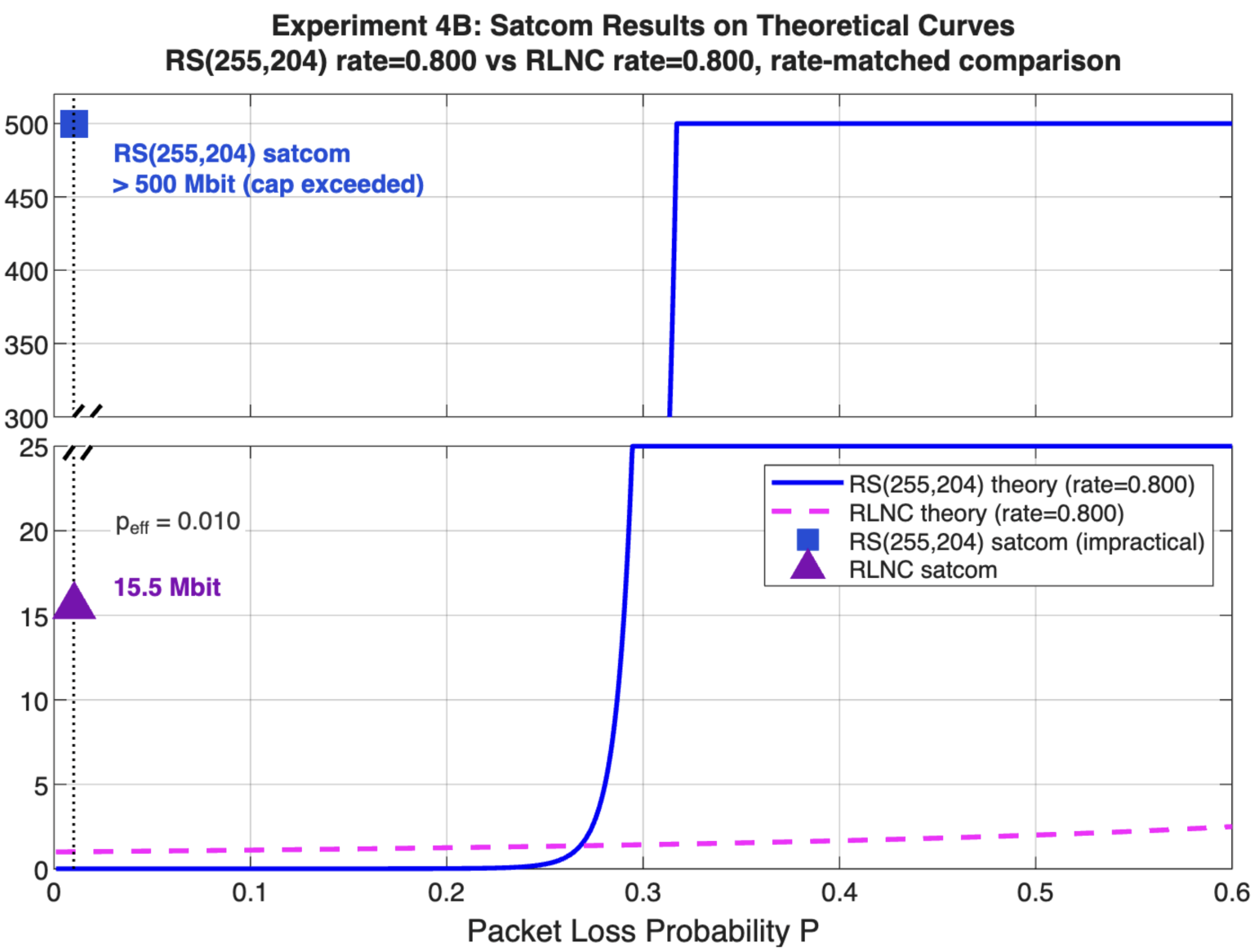}
\caption{Experiment 4B: Satcom results overlaid on theoretical curves at $\epsilon_G = 0.010$. The y-axis is broken to show both the RS$(255,204)$ satcom result (top panel, blue square), which exceeds the 500~Mbit transmission cap despite an open-link loss rate of only 1\%, and the RLNC satcom result (bottom panel, purple triangle) of 15.5~Mbit to retrieve all 1000 packets.  The RS theory curve rising steeply in the bottom panel illustrates the collapse regime into which the burst channel pushes RS$(255,204)$, while RLNC remains close to its theoretical prediction, confirming~\eqref{eq:ge_asymmetry}.}
\label{fig:exp4b}
\end{figure}

Fig.~\ref{fig:exp4b} provides the key quantitative comparison. At an open-link loss rate of $\epsilon_G = 0.010$, i.i.d.\ theory predicts RS$(255,204)$ should require only approximately 1~Mbit. Instead, the satcom simulation finds RS$(255,204)$ is entirely impractical, exceeding the 500~Mbit transmission cap. The RS theory curve rising steeply in the lower panel illustrates why: the burst structure of the satellite channel effectively pushes RS$(255,204)$ into the collapse regime around $p \approx 0.20$, even though the open-link loss rate is only 1\%. RLNC requires approximately 15.5~Mbit, sitting close to its theoretical prediction, confirming \eqref{eq:ge_asymmetry}: a system designer using $\epsilon_G$ alone to select between RS and RLNC would incorrectly choose RS. The burst penalty makes the average loss rate a misleading metric for code selection on satellite channels.

\subsection{Experiment 5: Burst Duration Sweep}
\label{sec:burst}

Experiment 5 isolates the effect of burst duration. The GE recovery probability $r$ is varied from 0.02 to 0.95, giving mean burst durations from 50 steps down to approximately 1 step, with $p_{\text{GE}} = 0.05$ and $\epsilon_G = 0.05$ fixed throughout.

\begin{figure}[tp]
\centering
\includegraphics[width=\columnwidth]{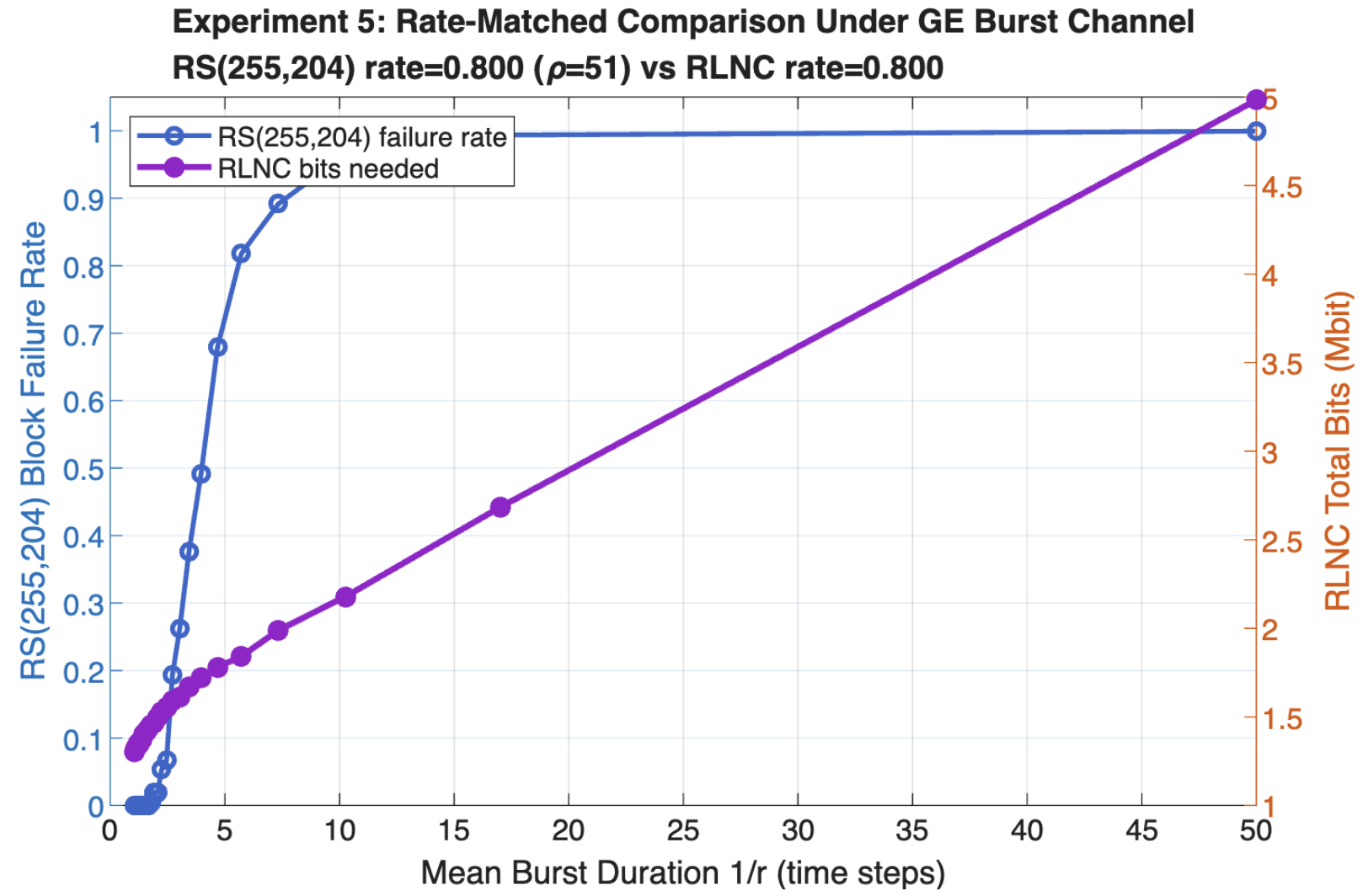}
\caption{Experiment 5: Rate-matched dual-axis comparison at code rate$=0.800$. RS$(255,204)$ failure rate (left axis, blue) rises sharply with burst duration, reaching near-100\% failure well before the satcom mean burst duration of 166~steps, while RLNC overhead (right axis, purple) grows linearly from 1.3~Mbit to 5.0~Mbit. The structural difference between RS and RLNC under bursty channels is not a rate effect but a fundamental property of block coding vs accumulation-based coding \cite{lun2006coding, biyikoglu2025modeling}.}
\label{fig:exp5}
\end{figure}

Fig.~\ref{fig:exp5} shows RS$(255,204)$ block failure rate and RLNC overhead on the same dual-axis plot at the matched rate of 0.800. With $\rho = 51$, RS$(255,204)$ failure rate remains near zero for short bursts, then rises sharply, reaching near-100\% failure well before the satcom scenario's mean burst duration of 166 steps, far beyond $\rho = 51$. The simulation rises faster than the pure geometric theory $(1-r)^{51}$ because within-state losses $\epsilon_G = 0.05$ compound with burst losses in the GE model \cite{biyikoglu2025modeling}. RLNC overhead, by contrast, grows linearly from approximately 1.3~Mbit to 5.0~Mbit as burst duration increases from 1 to 50 steps, a direct consequence of $\eta_{\text{RLNC}} \approx 1.25/(1-p)$: the generation continues accumulating across subsequent rounds so overhead grows proportionally rather than collapsing \cite{lun2006coding}.

This is the clearest demonstration of the paper's central finding: at identical code rates, RS collapses catastrophically under bursty channels while RLNC degrades gracefully. The difference is not attributable to rate, block length, or average loss; it is purely a structural consequence of RS's hard erasure capacity limit $\rho$ versus RLNC's accumulation property.

\subsection{Experiment 6: RS+RLNC Hybrid}
\label{sec:exp6}

Experiment 6 evaluates the concatenated RS+RLNC hybrid as a scheme providing simultaneous error and erasure correction. The RLNC outer code uses $K_{\text{gen}} = 16$, $N_{\text{coded}} = 20$, rate $0.800$ throughout. Three inner RS configurations from the same $N = 255$ family are tested \cite{vukobratovic2012unequal}.

On a pure erasure channel, inner RS provides no correction benefit since no bit errors are present: pure RLNC uses fewest bits at all loss rates, while the three hybrid configurations cluster near $2\times$ overhead relative to pure RLNC, consistent with their similar inner RS rates (0.749--0.875).

\begin{figure}[tp]
\centering
\includegraphics[width=\columnwidth]{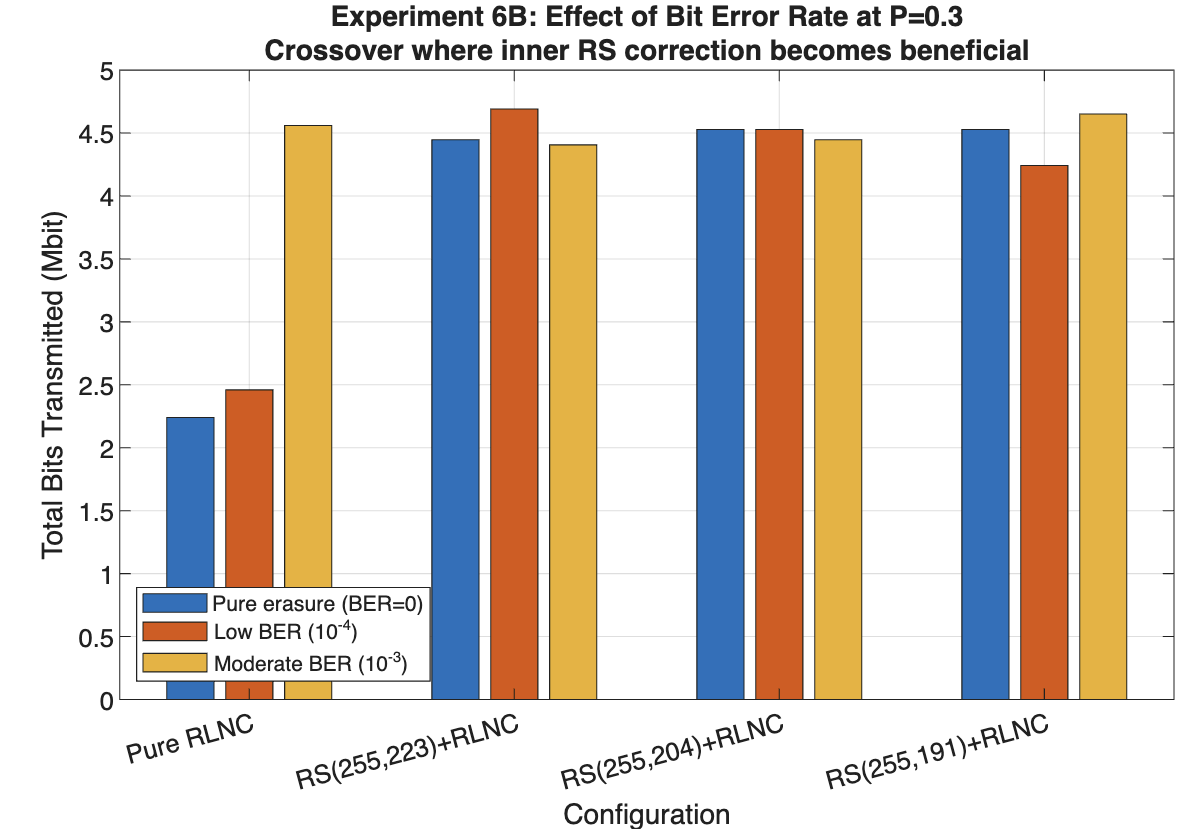}
\caption{Experiment 6B: Effect of bit error rate on each configuration at $P=0.3$. At BER$=0$ pure RLNC wins (2.24~Mbit vs 4.41--4.53~Mbit). At BER$=10^{-3}$ pure RLNC rises to 4.56~Mbit; hybrid results are mixed, with two of three inner RS rates falling slightly below pure RLNC and one slightly above \cite{vukobratovic2012unequal}.}
\label{fig:exp6b}
\end{figure}

Fig.~\ref{fig:exp6b} shows the BER sweep at $P = 0.3$. At BER$=0$, pure RLNC (2.24~Mbit) beats the hybrids (4.41--4.53~Mbit) due to unnecessary inner RS overhead. At BER$=10^{-3}$, pure RLNC rises to 4.56~Mbit; two hybrids fall slightly below this (RS$(255,223)$: 4.41~Mbit, RS$(255,204)$: 4.45~Mbit) while the third does not (RS$(255,191)$: 4.65~Mbit), indicating that any advantage from inner RS correction at this BER is modest and inconsistent across inner rates rather than a clean crossover.

\begin{figure}[tp]
\centering
\includegraphics[width=\columnwidth]{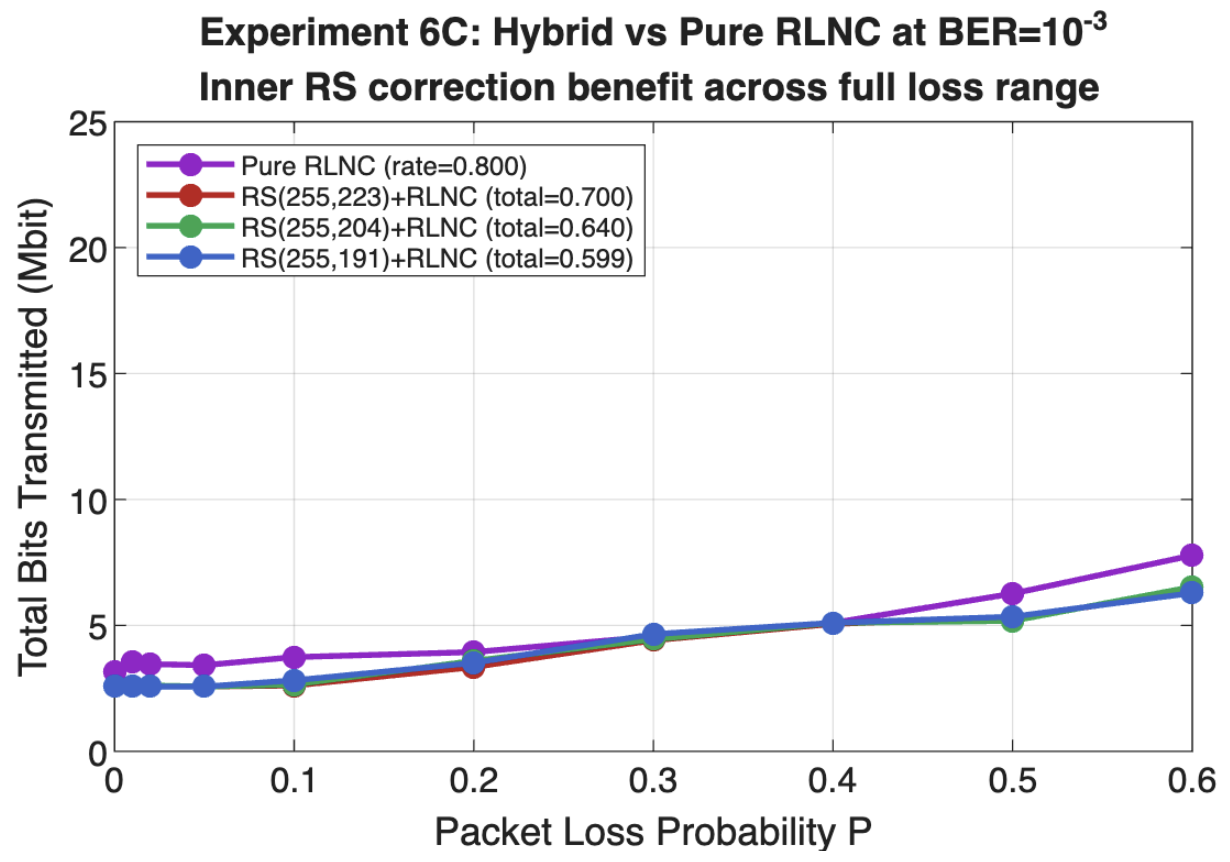}
\caption{Experiment 6C: Hybrid vs pure RLNC at BER$=10^{-3}$ across the full loss range. Pure RLNC (top, purple) is the worst performer at every loss rate. The three hybrid configurations perform similarly because their inner RS codes have similar correction capabilities at this BER
\cite{vukobratovic2012unequal}.}
\label{fig:exp6d}
\end{figure}

Fig.~\ref{fig:exp6d} shows the full fixed-$p$ sweep at BER$=10^{-3}$: pure RLNC is the worst performer throughout, while all three hybrids perform similarly (their $t$ values of 16--32 all suffice at this BER). At $p=0.6$, the hybrids require 6.28--6.53~Mbit versus 7.78~Mbit for pure RLNC, a 16--19\% reduction depending on inner RS rate.

\begin{figure}[tp]
\centering
\includegraphics[width=\columnwidth]{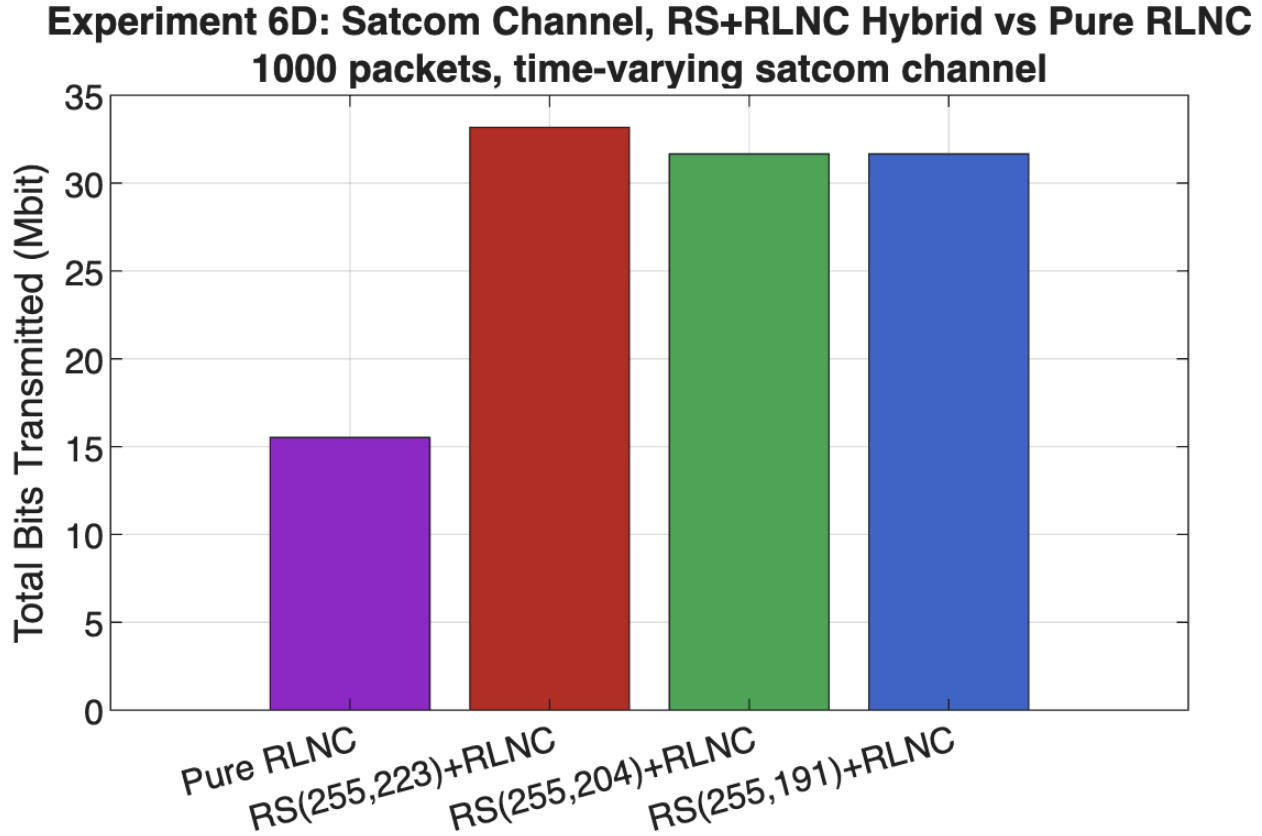}
\caption{Experiment 6D: Satcom channel results. Pure RLNC requires 15.5~Mbit. All three RS+RLNC hybrid configurations require approximately 32~Mbit, more than double pure RLNC, because the inner RS codes use $N_{\text{in}} = 255$ symbols per packet, doubling the bits per transmission. On a pure burst-erasure satcom channel, pure RLNC is the most efficient scheme.}
\label{fig:exp6c}
\end{figure}

Fig.~\ref{fig:exp6c} shows the satcom channel results: pure RLNC requires 15.5~Mbit to fetch all 1000 packets while all three hybrids require approximately 32~Mbit, more than double, since the inner RS codes use $N_{\text{in}} = 255$ symbols to encode what would otherwise be a 125-symbol packet, doubling bits per transmission and compounding with the burst structure. Pure RLNC is thus recommended for a pure burst-erasure channel, with the hybrid preferred only when bit errors and burst erasures both occur.

\section{Discussion}
\label{sec:discussion}

\subsection{The Crossover and Bridging Property: Summary}

The central finding of this paper is the existence and characterisation of the crossover loss rate $p^*$ and the bridging property of RLNC. By fixing both RS and RLNC at rate $0.800$, this paper isolates the structural difference between block-limited erasure correction and accumulation-based erasure correction \cite{lun2006coding} from the confounding effect of different rates. The results show that:

\begin{enumerate}
\item Below $p^* \approx 0.21$, RS$(255,204)$ and RLNC perform similarly, with RS$(255,204)$ slightly more efficient at very low loss due to its deterministic block structure \cite{chou2003practical}. 
\item Above $p^* \approx 0.21$, RLNC is decisively more efficient and the gap grows rapidly, purely because RS blocks fail with increasing frequency while RLNC generations continue accumulating across rounds. 
\item No RS code rate can be chosen to eliminate the crossover: every RS code has a finite $p^*$ above which it collapses, and the envelope of all three RS codes evaluated here collapses by $p \approx 0.27$. 
\item RLNC operates continuously through all three RS failure regions with a single fixed scheme \cite{lun2006coding}, confirming the bridging property analytically and experimentally. 
\end{enumerate}

The existence of a finite $p^*$ for every RS code and the absence of any corresponding crossover for RLNC is the fundamental asymmetry that distinguishes the two schemes, and it holds regardless of code rate, block length, or generation size.

These results are validated both on the memoryless fixed-$p$ channel (Experiment~3) and on the physically derived satcom channel (Experiment~4), where the burst structure pushes RS$(255,204)$ above its crossover despite $\epsilon_G$ of only 1\%, arguably the paper's most practically consequential finding.

\subsection{Practical Implications}

For practical satellite system design, three conclusions follow.

First, RLNC provides continuous, burst-resilient coverage across the full loss range, using estimated burst duration rather than average loss rate $\bar{\epsilon}$ as the relevant design metric: in deployments with mean burst duration reaching 166~minutes, as on real Pacific Island satellite links \cite{speidel2015network}, RLNC maintains practical overhead where a fixed block code would need $\rho > 166$ to match it. The bridging property (Section~\ref{sec:exp3}) further shows that a family of RLNC operating points can span the full range between any two block-code rates, illustrated in Fig.~\ref{fig:multi_rlnc}.

Second, RLNC offers this coverage as a single, structurally distinct solution \cite{koetter2003algebraic} rather than requiring a family of differently-parameterised codes. One RLNC configuration handles the full range of operating conditions that would otherwise require selecting among multiple fixed-rate alternatives, and continues to degrade gracefully well beyond the point where matching those alternatives would no longer be practical.

Third, the RS+RLNC hybrid provides dual protection against bit errors and burst erasures at the cost of approximately $2\times$ overhead relative to pure RLNC on the satcom channel. It becomes the recommended scheme when the satellite channel has significant residual bit errors in addition to burst erasures \cite{vukobratovic2012unequal}. HARQ using LDPC is the only other widely deployed scheme combining error and erasure correction \cite{landon2025rethinking}; RS+RLNC provides an alternative that inherits RLNC's burst resilience at the outer layer.

Fourth, the crossover loss rate $p^*$ is not a fixed property of RLNC but depends on $K_{\text{gen}}$ and $N_{\text{coded}}$. From~\eqref{eq:crossover_eq}, increasing $K_{\text{gen}}$ while holding the code rate constant shifts $p^*$ slightly lower, making RLNC competitive with RS at lower loss rates. This suggests a natural adaptive strategy: operate RS below $p^*$ and switch to RLNC above $p^*$, using the estimated burst duration $1/\hat{r}$ as the switching signal rather than the average loss rate $\bar{\epsilon}$. Such a scheme would achieve near-optimal efficiency across the full operating range of the satellite channel, combining the bandwidth efficiency of RS at low loss with the burst resilience of RLNC under the bursty satcom channel. This represents the most immediate direction for future work, and connects directly to the adaptive unequal error protection strategies studied in \cite{vukobratovic2012unequal}.

\subsection{Limitations}
\label{sec:limitations}

The fixed-$p$ sweep uses an independent erasure model that does not capture the real channel's burst correlation; the GE model itself is a two-state simplification, and more accurate finite-state Markov models better fit real LEO traces \cite{fontan2022review, lopezsalamanca2020cubesat} but require measurement data not available here. The satcom results (Experiment~4) derive from a single orbital configuration (10{,}000~km semi-major axis, MEO) and one ground-station pair; LEO constellations, the focus of most 6G-NTN discussion, would produce shorter, more frequent outage windows requiring separate characterisation.

\section{Conclusion and Future Work}
\label{sec:conclusion}

This paper presented a mathematical and simulation-based comparison of Reed-Solomon coding and Random Linear Network Coding at matched code rates in a three-hop satellite communication channel, establishing an analytical characterisation of the crossover loss rate $p^*$ and the bridging property of RLNC. A physically grounded packet loss model was derived from orbital geometry and link budget calculations, and a Gilbert-Elliott channel model was fitted to the resulting burst loss structure. Closed-form expressions for the decoding success probability, overhead ratio, crossover loss rate, and burst failure probability were derived and validated against MATLAB simulation across six experiments.

Three findings stand out. RLNC and RS$(255,204)$ perform similarly below $p^* \approx 0.21$, above which RLNC is decisively more efficient purely due to burst resilience, not rate advantage. The bridging property means a single RLNC configuration \cite{lun2006coding} replaces the entire family of RS code rates across all loss regions. Finally, under the derived satcom channel (mean burst duration 166~minutes), RS$(255,204)$ is impractical despite only 1\% average loss, confirming burst structure, not $\bar{\epsilon}$, as the critical factor for code selection.

Future work will investigate the crossover behaviour for larger generation sizes \cite{luby2002lt, shokrollahi2006raptor}, adaptive switching between RS and RLNC based on estimated burst duration \cite{vukobratovic2012unequal}, sliding window RLNC for reduced decoding delay \cite{gabriel2020sliding}, and validation against real LEO satellite packet traces \cite{fontan2022review} using more accurate channel models such as the Markovian Arrival Process.

\bibliographystyle{IEEEtran}
\bibliography{citations}

\end{document}